\pdfoutput=1
\input{aipcheck}
\documentclass[final
  ]
  {aipproc}
\usepackage{graphicx}
\usepackage{subfigure}
\layoutstyle{6x9}

\begin{document}

\title{Dihadron Fragmentation Functions in the NJL-jet model}

\author{A. Casey, A. W. Thomas, H. H. Matevosyan}{
  address={CoEPP and CSSM, School of Chemistry and Physics, University of Adelaide, 5005}
}

\begin{abstract}
The NJL-jet model provides a framework for calculating fragmentation functions without introducing ad hoc parameters. Here the NJL-jet model is extended to investigate dihadron fragmentation functions.
\end{abstract}

\maketitle

\vspace{-0.8cm}
\section{Introduction}

Deep inelastic scattering (DIS) has proven to be an invaluable source of information about the structure of the nucleon\cite{weisethomasbook}, providing insight into parton distribution functions. Semi-inclusive deep-inelastic scattering (SIDIS) has extended our understanding further, allowing further access to the transverse structure of nucleons.

With a deeper understanding of the fragmentation functions, these studies will continue to provide us with valuable information on nucleon structure. Fragmentation functions are an important theoretical tool in the investigation of scattering reactions. This has led to the development and study of the NJL-jet model \cite{LB1,kaonffnjl,kaonfrag}. The NJL-jet model builds on the Field-Feynman Quark-jet model (FFQJM)\cite{FF2}, by using an effective chiral quark model framework in which calculations of both quark distribution and fragmentation functions can be performed without introducing ad hoc parameters. Pion fragmentation functions in the NJL-jet model were calculated in Ref. \cite{LB1}. The NJL-jet model was then extended to include strange quark contributions, and kaon fragmentation functions were obtained in Ref. \cite{kaonffnjl} and in Ref. \cite{kaonfrag} the fragmentation functions to vector meson and nucleon-antinucleons were calculated.

\vspace{-0.3cm}
\section{Fragmentation Functions and the NJL-jet model}

In Fig. \ref{fig:quarkcascade}, it is shown how a quark can produce a cascade of hadrons, producing what is interpreted as jet events in Deep-Inelastic Scattering (DIS). It is important to note that within the model the emitted hadrons do not interact with the other hadrons produced in the quark jet. An integral equation for the quark cascade process shown in Fig. \ref{fig:quarkcascade} is derived in the NJL-jet model of Ref. \cite{LB1}. The integral equation for the total fragmentation function is:
\begin{figure}[h!]
\includegraphics[width=0.55\textwidth]{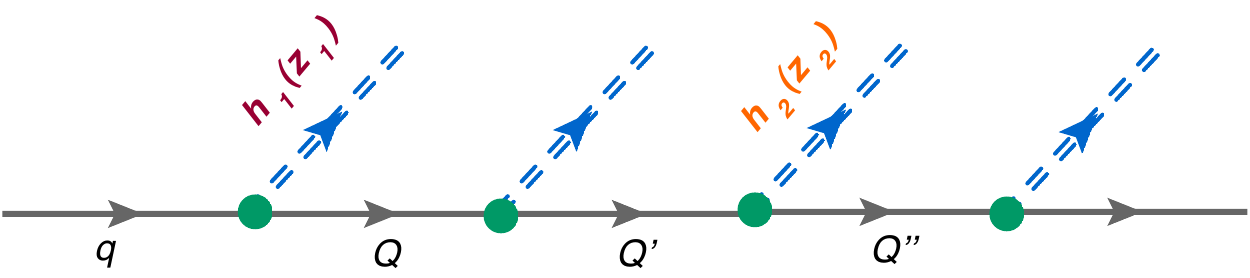}
\caption{Quark Cascade} 
\label{fig:quarkcascade}
\end{figure}
\vspace{-0.3cm}
\begin{equation}
D^m_q(z)dz=\hat{d}^m_q(z)dz+\sum_{Q}\int^{1}_{z}\frac{dy}{y}\hat{d}^Q_q(\frac{z}{y})D
^m_Q(y)dz,\quad\quad\hat{d}^Q_q(z)=\hat{d}^m_q(1-z)|_{m=q\bar{Q}}
\end{equation}

The first term on the right hand side is the driving function and represents the probability of creating a meson $m$ carrying momentum fraction $z$ to $z+dz$ from the initial quark(i.e. $h_1=m$). The second term represents the probability of creating the meson, $m$, further in the quark decay chain(i.e. $h_2=m$). The method used to solve this equation approximates integrals as sums over discretized values of $z$, so that $D^m_q(z)$ and $\hat{d}^m_q(z)$ can be written as vectors and the integral term (without $D^m_q(z)$) can be written as a matrix, where the elements are their values at $z=z_i$. In index form the elements of the vectors can be written as $D^m_{q\, i}=f_i+g_{ij}D^m_{q\, j}$. Writing in vector form and rearranging to solve for $\vec{D}^m_q$, we obtain $\vec{D}^m_q = (I-g)^{-1}\vec{f}$, where $f$ is taken to be the driving function and $g$ is the matrix form of the integrand without the fragmentation function.
\vspace{-0.3cm}
\section{Dihadron Fragmentation Functions}

The extension to dihadron fragmentation functions in the NJL-jet model is now considered. Dihadron fragmentation functions $D^{h_1,h_2}_q(z_1,z_2)$ correspond to the probability of observing two hadrons, $h_1$ and $h_2$, with light-cone momentum fractions $z_1$ and $z_2$, respectively, fragmenting from the initial quark $q$. An illustration of how a quark cascade can produce two observed hadrons in the NJL-jet model is shown in Fig. \ref{fig:quarkcascade}, where $h_1$ and $h_2$ are the observed hadrons.

The integral equation for the dihadron fragmentation function $D^{h_1,h_2}_q(z_1,z_2)$ constructed from Field and Feynman's Eqs (2.43a)-(2.43d) of Ref. \cite{FF2} is:
\begin{eqnarray}
\label{DFF2}
D^{h_1,h_2}_q(z_1,z_2) & = & \hat{d}^{h_1}_q(z_1)\frac{D^{h_2}_{q_1}(\frac{z_2}{1-z_1})}{1-z_1}+\hat{d}^{h_2}_q(z_2)\frac{D^{h_1}_{q_2}(\frac{z_1}{1-z_2})}{1-z_2}\\ \nonumber
 & &+\sum_Q\int^{1}_{z_1+z_2}\frac{d\eta}{\eta^2}\hat{d}^Q_q(\eta)D^{h_1,h_2}_Q(\frac{z_1}{\eta},\frac{z_2}{\eta}), \quad q\rightarrow h_1+q_1;\quad q\rightarrow h_2+q_2,
\end{eqnarray}

where $\hat{d}^h_q(z)$ and $\hat{d}^Q_q(\eta)$ are the elementary splitting functions of the quark $q$ to the corresponding hadron $h$ and quark $Q$. The first term corresponds to the probability of producing hadron $h_1$ from quark $q$ at the first step in the cascade, followed by hadron $h_2$ either directly afterwards or further down the cascade chain. Similar for the second term with hadrons $h_1$ and $h_2$ switched in order. The third term corresponds to the probability of both hadrons being produced after the first step of the cascade.

Collecting the first two terms of Eq.(\ref{DFF2}) together, a term analogous to the driving function of the single hadron fragmentation function is found for the dihadron fragmentation function. Due to the discretization method used to solve the equation, the division by $\eta$ causes the values of $D^{h_1,h_2}_Q(\frac{z_1}{\eta},\frac{z_2}{\eta})$ obtained to not correspond to the discrete values in the chosen region for $z_1$ and $z_2$. To fix this issue, the values of $z_1$ and $z_2$ were discretized in the region [0,1] and new variables were set as $\xi_1=\frac{z_1}{\eta}$ and $\xi_2=\frac{z_2}{\eta}$. Values for  $\xi_1$ and $\xi_2$ were discretized in the same way as $z_1$ and $z_2$ in the region [0,1]. Rewriting the integral part of Eq.(\ref{DFF2}) in terms of these new variables, the dihadron fragmentation function becomes:
\begin{eqnarray}
D^{h_1,h_2}_q(z_1,z_2) & = & \hat{d}^{h_1}_q(z_1)\frac{D^{h_2}_{q_1}(\frac{z_2}{1-z_1})}{1-z_1}+\hat{d}^{h_2}_q(z_2)\frac{D^{h_1}_{q_2}(\frac{z_1}{1-z_2})}{1-z_2}\\ \nonumber
& & +\sum_Q\int^{\frac{z_1}{z_1+z_2}}_{z_1}d\xi_1\int^{\frac{z_2}{z_1+z_2}}_{z_2}d\xi_2\delta(z_2\xi_1-z_1\xi_2)\hat{d}^Q_q(z_1/\xi_1)D^{h_1,h_2}_Q(\xi_1,\xi_2)\\ \nonumber
& &
\end{eqnarray}

\begin{figure}[h!]
\hspace{0.1cm}
\includegraphics[trim=0.2cm 0.2cm 1.9cm 0.5cm,clip,width=0.49\textwidth]{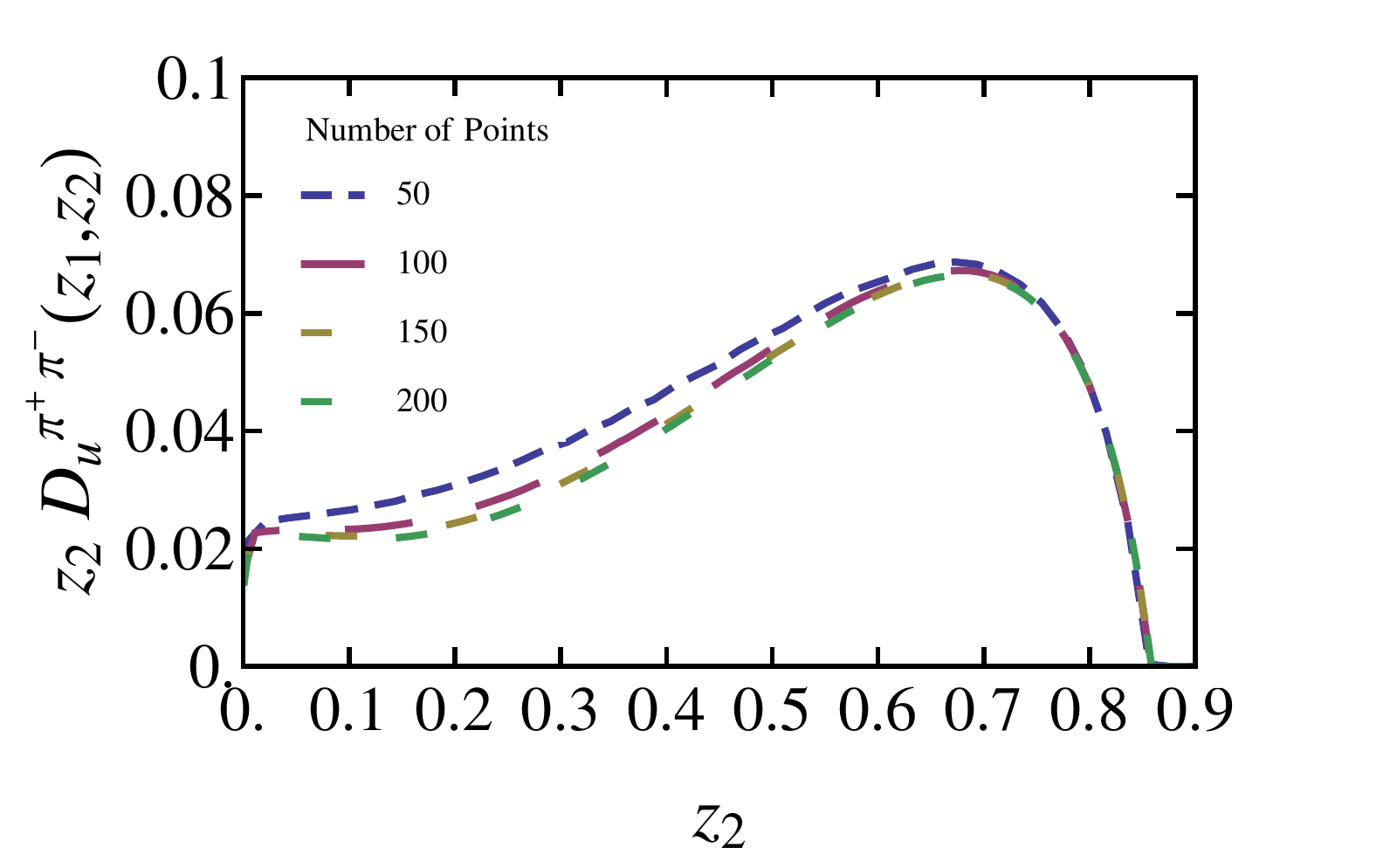}
\hspace{0.1cm}
\includegraphics[trim=0.2cm 0.2cm 2.2cm 0.5cm,clip,width=0.49\textwidth]{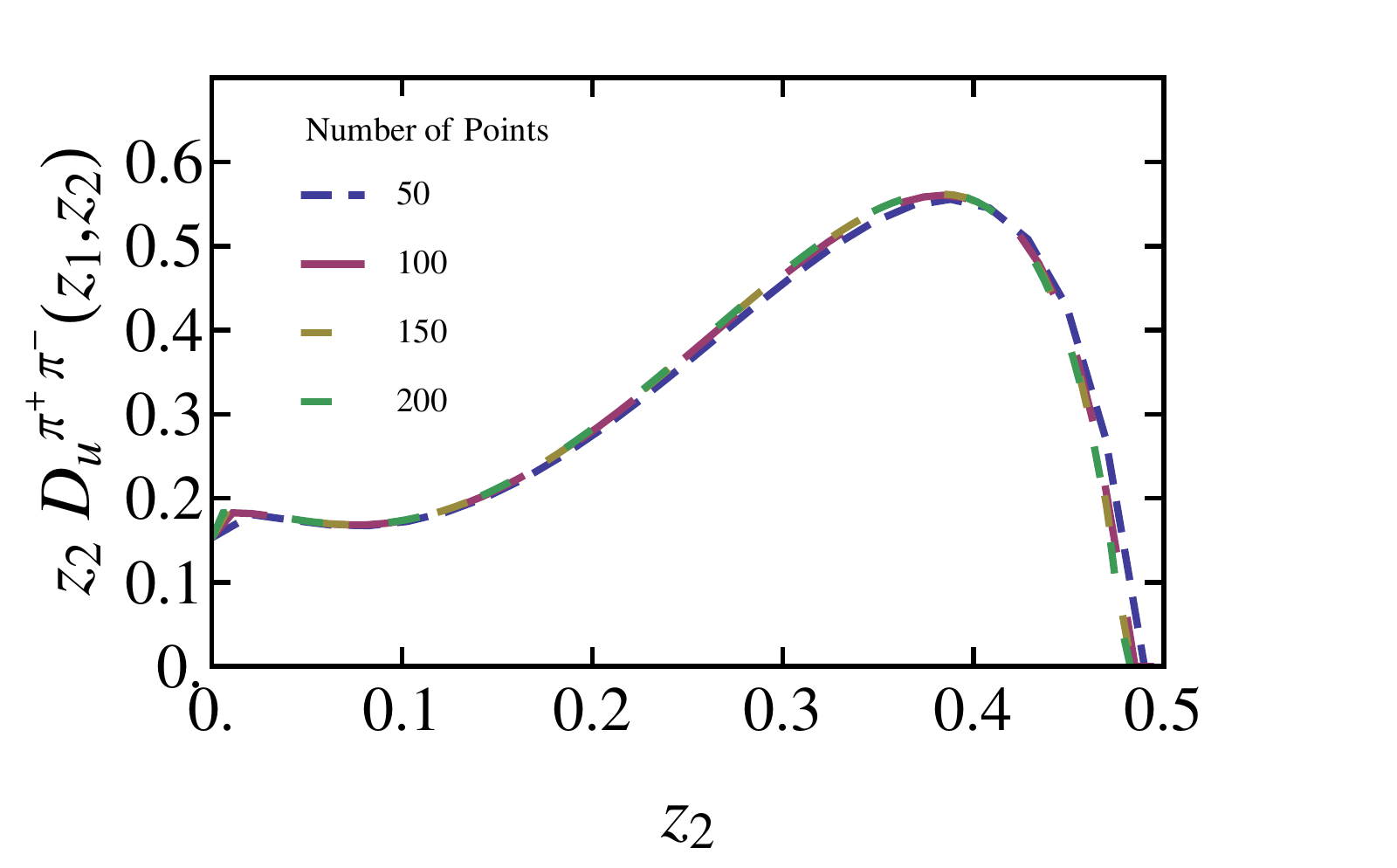}
\caption{Dihadron fragmentation functions at the model scale for $h_1=\pi^+$, $h_2=\pi^-$ for varying number of $z$ points at (left) $z_1=0.1$ and (right) $z_1=0.5$}
\label{fig:plotcompare}
\end{figure}

Mathematica was used to solve for the single hadron and dihadron fragmentation functions. The number of points used for the dihadron cases was 200, while the single hadron case was 500 points. Fig. \ref{fig:plotcompare} shows the results for the dihadron fragmentation function $D^{\pi^+ \pi^-}_u(z_1,z_2)$ solved for 50, 100, 150 and 200 points at $z_1=0.1$(left) and $z_1=0.5$(right). For the $z_1=0.5$ case there is very good agreement for each choice of the number of points. For the $z_1=0.1$ case, sufficient convergence of the function is obtained by 200 points, and so this is the number of points chosen to be used. To solve for the delta function in the integral term the values of $z_1$ and $z_2$ are looped over first, and then $\xi_1$ is looped over($z_1$ and $z_2$ share the same range of values as $\xi_1$ and $\xi_2$, and are discretized in the same way). This leaves $\xi_2$ with a value obtained from the delta function that may not be one of the discretised values. The values of the DFFs at $\xi_2$ were obtained using linear interpolation from neighbouring discrete values.

\section{Results}
\label{sec:dfcontri}
\begin{center}
\begin{figure}[h!]
\includegraphics[trim=0.3cm 0.0cm 1.9cm 0.4cm,clip,width=0.49\textwidth]{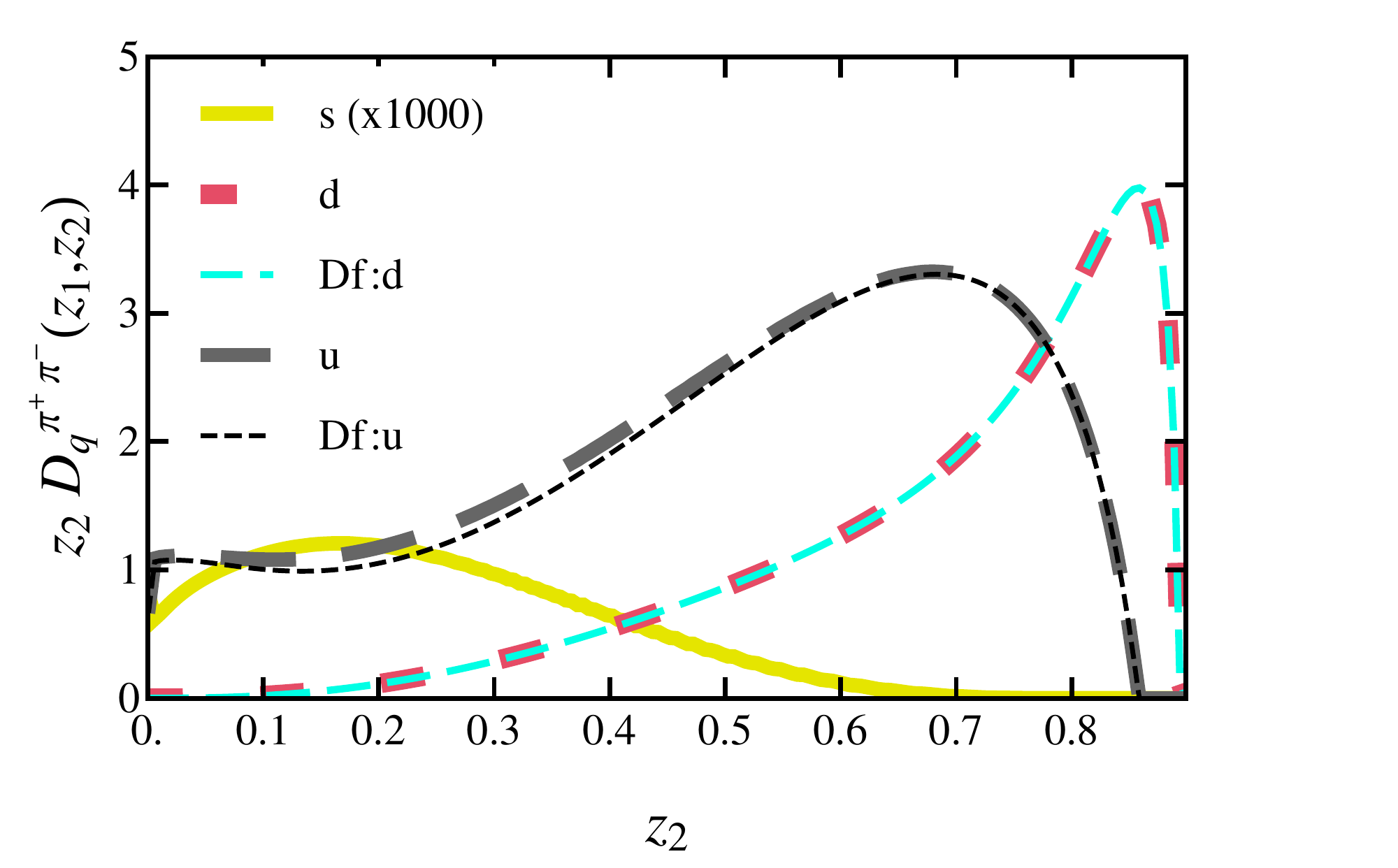}
\includegraphics[trim=0.3cm 0.0cm 1.9cm 0.4cm,clip,width=0.49\textwidth]{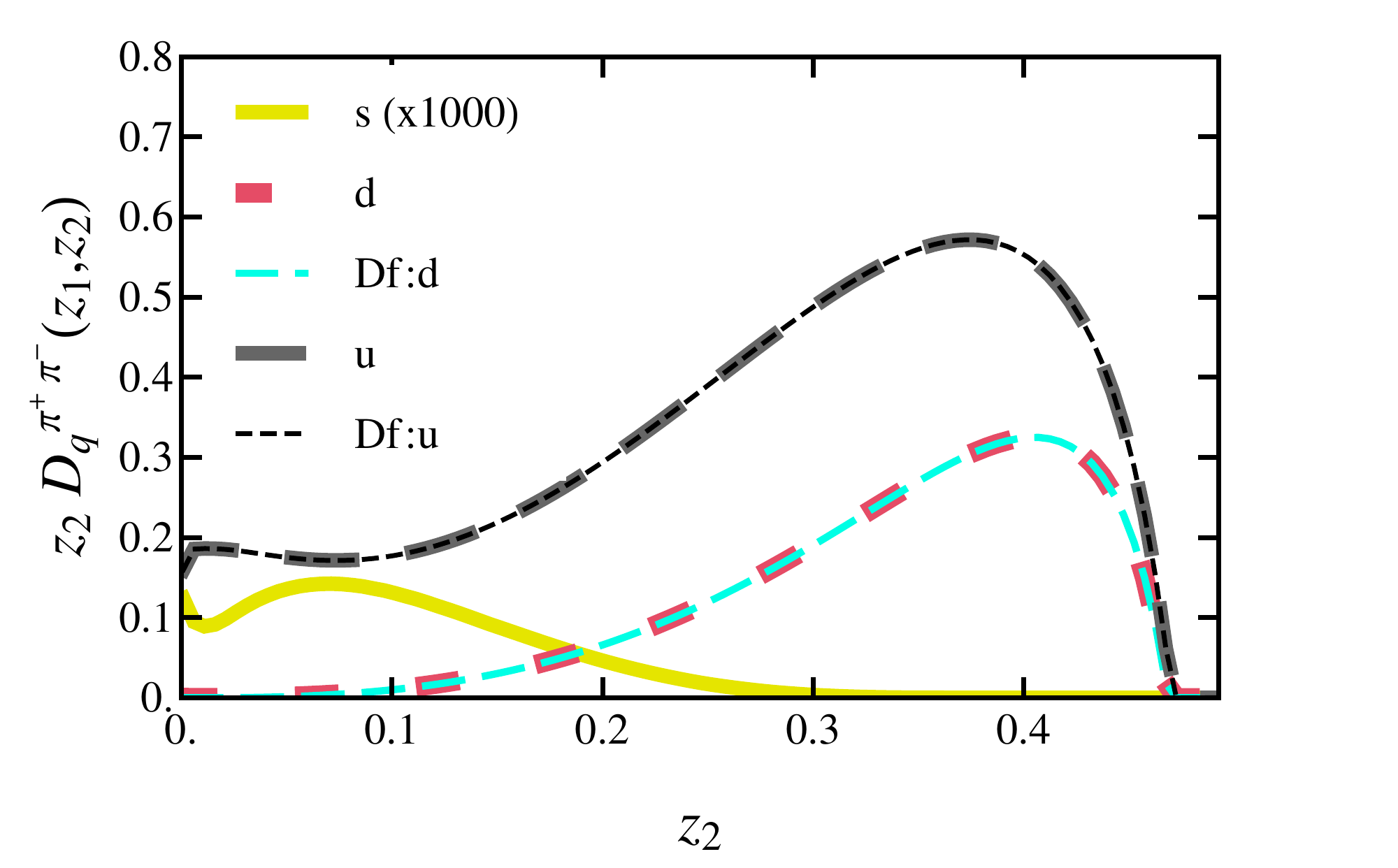}
\caption{Dihadron fragmentation functions at the model scale for $h_1=\pi^+, h_2=\pi^-$ at (left) $z_1=0.1$ and (right) $z_1=0.5$}
\label{fig:drivfunc}
\end{figure}
\end{center}

Using the data obtained from solving for the dihadron fragmentation functions, it is possible to compare the solutions ($z_2\, D^{h_1h_2}_q$) to the driving function (the first two terms of Eq. \ref{DFF2}). In Fig. \ref{fig:drivfunc} it can be seen that the driving functions contribute almost all of the DFF. The figure on the left shows that for an initial up quark, there is a slight difference between the DFF and the driving function at low values of $z_2$ for a low value of $z_1$(=0.1). The strange quark has a zero value driving function, and therefore is completely made up of the integral term. The difference between the DFFs and the driving functions for each case is of the same order of magnitude as the strange quark DFF.  This difference is only visible for the initial up quark case when the driving function becomes small, which occurs at low values of $z_1$.

\vspace{-1.3cm}
\section{Conclusions}
In this document, results have been presented for dihadron fragmentations calculated within the NJL-jet model. The difference between the driving functions and the full DFFs was shown to become apparent only at low $z_1$. The integral term for each case is of the same order of magnitude, though they are not equal. The difference is visible for the initial up quark case only where the driving function becomes small.
\vspace{-0.3cm}
\begin{theacknowledgments}
This work was supported by the Australian Research Council(Grant No. FL0992247) and by the University of Adelaide.
\end{theacknowledgments}

\vspace{-0.3cm}
\bibliographystyle{aipproc}

\end{document}